\documentclass[a4paper]{article}

\usepackage[english]{babel}
\usepackage[utf8]{inputenc}
\usepackage{amsmath}
\usepackage{graphicx}
\usepackage[colorinlistoftodos]{todonotes}
\usepackage{subcaption}
\usepackage{authblk}

\newcommand{\argmax}{\operatornamewithlimits{argmax}}

\title{Automatic Neuron Type Identification by Neurite Localization in the {\it Drosophila} Medulla}

\author{Ting Zhao}
\author{Stephen M Plaza}
\affil{Howard Hughes Medical Institute, Janelia Farm Research Campus Ashburn, VA 20147}

\date{\vspace{-5ex}}

\begin{document}
\maketitle

\begin{abstract}
Mapping the connectivity of neurons in the brain (i.e., connectomics) is a challenging problem due to both the number of connections in even the smallest organisms and the nanometer resolution required to resolve them.  Because of this, previous connectomes contain only hundreds of neurons, such as in the C.elegans connectome.  Recent technological advances will unlock the mysteries of increasingly large connectomes (or partial connectomes).  However, the value of these maps is limited by our ability to reason with this data and understand any underlying motifs.  To aid connectome analysis,
we introduce algorithms to cluster similarly-shaped neurons,
where 3D neuronal shapes are represented as skeletons. 
In particular, we propose a novel location-sensitive clustering
algorithm.  We show clustering results on neurons reconstructed from the
Drosophila medulla that show high-accuracy. 
\end{abstract}

\section{Introduction}
Advances in producing large electron microscopic (EM)
datasets enables the identification of detailed neuronal pathways and synapses.
However, this resolution requires the collection
of a lot of raw data.  This data must be processed automatically and manually requiring
time-consuming connectome reconstruction \cite{plaza2014toward}.
The resulting connectome defines a graph of neurons
and their connections.  Often a connectome contains information on the characteristics or type
of each neuron, the strength of the connections, and the shape of the neuron.  The desired information will depend on the application.  For example, when modeling the electrical behavior of a neuron circuit, the shape and volume of the neuron are important.  

This paper aims to both improve laborious connectome reconstruction and provide a better mechanism to explore and reason with the resulting connectome.  The key contribution is a novel algorithm to cluster similar cells shapes.  Because of structural repetition in various neuropil, such clustering could provide a mechanism to validate reconstruction correctness
by measuring against similar neurons.   Also, clustering could provide a means to evaluation if each cluster represents a functional neuron cell type class, as we now discuss.

\begin{figure}
\centering
\includegraphics[width=0.8\textwidth, viewport=0 130 756 453, clip=true]{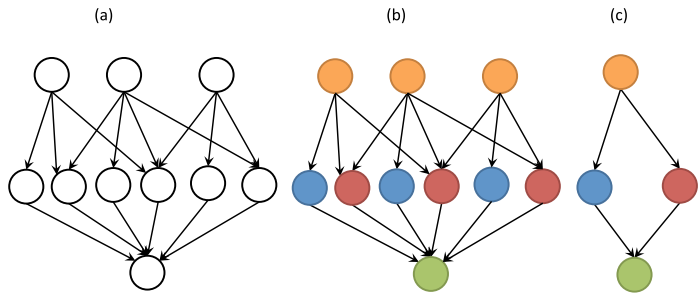}
\caption{\label{fig:connectome} Cell type information helps deciphering a neuron network (a). After label each neuron with a type, as shown in unique colors in (b), we can reduce the network into a simple 4-node graph (c), which reveals two pathways of information flow.}
\end{figure}

In spite of existing debates, one widely accepted definition of the neuronal cell type is a group of neurons that have similar functions \cite{masland2004neuronal}. This definition is not only necessary for communicating biology, but also helpful for analyzing neuronal circuits or a connectome more efficiently. An example of this
usefulness is our reconstruction of the 
motion detection circuit in the medulla \cite{takemura2013visual}. To identify this circuit, we first labeled the cell types
for all 379 neurons reconstructed, and then built a graph where each node is a neuron type and each edge indicates the connection strength between two types.  This enabled the identification of several strongly connected sub-circuits, including one involving L1, Tm3, Mi1 and T4 neurons. Further analysis on the Mi1 and Tm3 paths showed
a subtle offset in receptive fields and 
convergence to a common cell type, revealing
the motion circuit.
Figure \ref{fig:connectome} shows a connectivity graph with and without types annotated.  Without
annotations, one can discern data flow, but one cannot discern two different cell types in the middle layer.  Having nodes with different
characteristics, like delay, is an ingredient in motion detector models like Hassenstein-Reichardt \cite{reichardt}.

While the fundamental definition of neuronal
type is by function, the morphology or shape
of a neuron can act as a proxy.  In
\cite{takemura2013visual}, cell types were
defined by shape and corroborated with light
data that reflect genetic differences between
cell types \cite{jenett2012gal4}.  
Neuron shape is an accessible feature in
EM datasets and is an 
informative indicator of how the neuron connects to others in the brain network. The shape is even more informative when we have a reference framework describing where the neuron is located.
Shape and location act as a powerful mechanism
to distinguish neurons as in layered structures
like the fly medulla or mammal neocortex.

In this paper, we introduce a novel algorithm
that exploits the locality of neuronal arborisations
to cluster common cell types together.
The major deficiencies of previous cell clustering efforts, either based on shape description features \cite{tsiola2003quantitative, schierwagen2010cluster, lu2014quantitative} or shape matching \cite{basu2011path2path, heumann2009tree}, is that they do not consider the problem within a geometrical reference.
This is likely due to the limitation of most public data used for testing, which are obtained from different brains and where information
to register them to a common, standard brain, is missing.  In this paper, we consider a dataset
extracted from a single brain and whose location
is known within a standard model.
While the authors of \cite{masland2004neuronal}
believe that neuron shape reflects its function
by how it connects to others, the location
of neuron branching should be a more direct,
and thus more reliable, indicator of neuronal
function as compared to location-independent
shape features.  We can test this hypothesis
by identifying neuron types based on where
their branches are located in a common
reference frame.  We also note this might
be advantageous in matching partial shapes, a
reality due to datasets and reconstructions that
do not encompass the entire brain.

Automatic cell clustering contributes to EM
connectomics in several ways.
First, automatic cell clustering can boost the speed
of reconstructing a connectome.
For instance, in previous work \cite{takemura2013visual}, we 
identified all cell types manually, a tedious task requiring expertise on the specific medulla organization.  In addition, automatic clustering can help find mistakes or missing branches in
the reconstruction process.  Finally, the objective nature of automatic methods is a more convincing way to confirm that neuron shape clusters are indeed natural phenomena rather than human brain illusion.

The paper is organized as follows.  We first
review background techniques in shape clustering.
Then, we introduce our neuron skeletonization
strategy and cell clustering algorithm.  We provide results on a Drosophila medulla dataset
from \cite{takemura2013visual} in the experiments.  We conclude with a discussion
on the generalizability and relevance of our approaches.


\section{Background}
There are various automatic cell type clustering methods reported in the literature based on cell shapes. One common way of clustering neurons is converting each neuron into a point in a high-dimensional feature space and performing clustering in the new space \cite{tsiola2003quantitative, schierwagen2010cluster, schierwagen2010cluster, lu2014quantitative}.
Those methods usually require a large set of features because of the complexity of neuron structures. For example, the widely used L-measure has about one hundred features \cite{scorcioni2008measure}, and Lu et al. added 25 more features to the L-measure for their method \cite{lu2014quantitative}. More features do not only mean high computational cost, but also increase the risk of over-fitting.


An alternative way is to match neurons based on the topology, which is similar to matching a tree, and measure the similarity between two neurons by their matching degree. Complete unconstrained tree matching is an NP-complete \cite{zhang1992editing} problem and we should either search for a sub-optimal solution or define a reference point to reduce the search space. For example, Heumann and Wittum adopted the tree-edit-distance metric to  match the compartments between two neurons \cite{heumann2009tree}. Basu et al. decomposed a neuron structure into a collection of overlapped paths and then register the path \cite{basu2011path2path}. The matching is to find the optimal match between two collections of paths. Those methods are usually very sensitive to root initialization. On the other hand, despite that numerous matching methods are proposed, there is limited work about clustering neurons with the similarity matrix generated from matching without explicit feature space definition. One piece of related work is building a dendrogram from the tree-edit-distance \cite{heumann2009tree}, but the authors did not analyze how many neuron clusters could be derived in an unsupervised way.
\section{Methods}

\subsection{Shape Comparison}

\subsubsection{A General Form of Location-Based Shape Comparison}

Comparing two neuron shapes is the same as computing the morphological similarity between two neurons, which can be defined in many ways. In our location-based method, the underlying assumption is that the neurons are more similar if they
 have more similar branch distributions at the same location or locations performing the same function. Thus, given two neurons $\mathcal{N}_1$ and $\mathcal{N}_2$ and their branch densities  $D_1(\mathbf{x})$ and  $D_2(\mathbf{x})$ at location $\mathbf{x}$, we define the similarity between the two neurons as

\begin{equation}
S(\mathcal{N}_1, \mathcal{N}_2) \propto \max_{W} \left( \int_{\mathbf{x}}  S_D \left(D_1(W(\mathbf{x})), D_2(\mathbf{x})\right)C\left(\mathbf{x}, W(\mathbf{x})\right)- \mathcal{P}(|W'(\mathbf{x})|) d\mathbf{x}\right)
\label{eq:matchg}
\end{equation}
where $S_D$ is the similarity between the branch densities,  $C(\mathbf{x}_1, \mathbf{x}_2) \le 1$  is the offset cost of two locations $\mathbf{x}_1$ and $\mathbf{x}_2$ in the same brain region.$W$ is the warping function to tolerate biological variations or registration uncertainties and $|W'(\mathbf{x})|$ denotes the local size change after warping, which gives a penalty $\mathcal{P} \ge 0$  to favor more size-invariant warping. Defining $S(\mathcal{N}_1, \mathcal{N}_2)$  in a proportional form reserves the possibility of improving comparison by adding optional factors, such as normalization or the similarity of global morphological features.

\subsubsection{The Specific Case of Comparing Medulla Neurons}

\vspace{2mm}
\noindent
{\bf Branch Density Matching}
\vspace{1mm}

\begin{figure}
  \centering
  \includegraphics[width=0.9\textwidth]{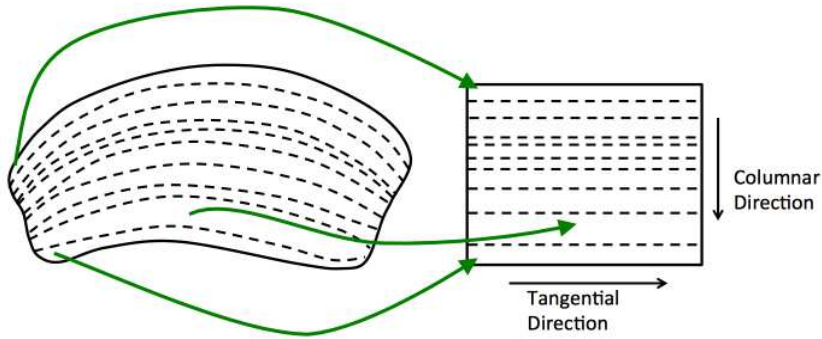}
  \caption{\label{fig:medulla} The medulla region can be mapped to a tow-dimensional rectangle based on its layer definition (dashed lines) given by \cite{fischbach1989optic}, with the sides of the rectangle being parallel to two principal directions, the columnar direction and the tangential direction.}
\end{figure}

The optic medulla is well known to be separated into ten layers of structures corresponding to its information flow, which is parallel to the orientation of the periodic column structure in the medulla \cite{fischbach1989optic}. We refer to this direction of information flow as the columnar direction and its perpendicular direction as the tangential direction. With these two directions, we can map the medullar into a 2D rectangle, as in Figure \ref{fig:medulla}. Since many medulla cells are identifiable based on the columnar position of their arborisations, as in  \cite{fischbach1989optic}, we can further simplify the problem by mapping the region into a 1D segment.

As a result,  we can define a quantitative measurement of the branch density as

\begin{equation}
D(z) = \sum_i T(g_i, z)
\end{equation}
where $g_i$is a neuron segment and $T(g_i, z)$ is the number of intersections between $g_i$ and the tangential plane at $z$. To facilitate implementation, we need to discretize $z$ into $N$ planes $(z_1, z_2, ..., z_N)$ , so the density similarity between plane $z_i$ and $z_j$ can be denoted as $s(i,j)$.  Following Eq. \ref{eq:matchg} ,we have the objective function

\begin{equation}
  \begin{array}{ccc}
    \sum_{k=1}^{K} s(m_k, n_k)c(z_{m_k},z_{n_k}) - P
  \end{array}
\end{equation}

where $c(z_1,z_2)$ is the cost function of $Z$ difference and $P$ is the gap penalty, corresponding to the terms of location offset cost and size penalty respectively in Eq. \ref{eq:matchg}. $c(z_1, z_2)$ helps tolerate the shift of the planes caused by biological variation, sample preparation, or the general inexactness due to the fixed coordinate system. In our implementation, we let 

\begin{equation}
c(z_1,z_2) = \frac{1}{t|z_1-z_2| + \lambda}
\end{equation}
where $t$ is the tolerance factor controlling how much the $Z$ offset can be tolerated and $\lambda$ is the regularization factor to guarantee the numerical stability of division.

The optimization attempts to find $\{(m_1, n_1), (m_2, n_2), \cdots, (m_t, n_t)\}$  that maximizes the objective function with the constraint that $m_i \le m_j$ and $n_i \le n_j$ if $i < j$. This can be solved efficiently by dynamic programming.

After obtaining the similarity matrix
\begin{equation}
S_{ij} = S(\mathcal{N}_i, \mathcal{N}_j)
\end{equation}

we normalized it as
\begin{equation}
\hat{S_{ij}} = \frac{S_{ij}}{\max(S_{ii}, S_{jj})}
\end{equation}

The normalization makes the similarity less sensitive to size so that a neuron is not biased to types with larger sizes.

\vspace{2mm}
\noindent
{\bf Tangential calibration}
\vspace{1mm}

\begin{figure}
\centering
\includegraphics[width=0.8\textwidth]{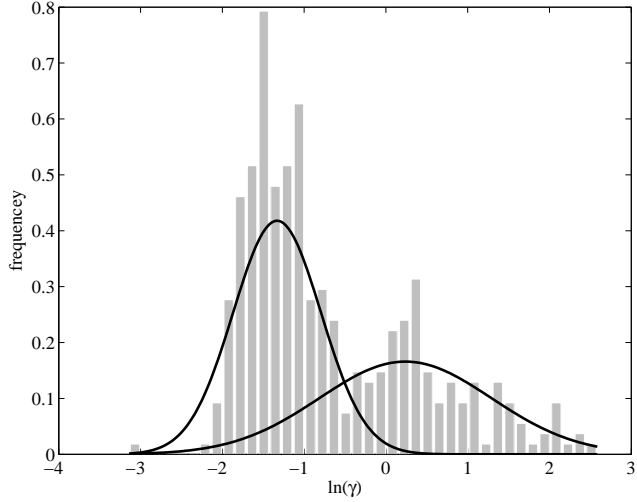}
\caption{\label{fig:lvratio} The mixture distributions of the logarithm  of tangetial-columnar span ratio (denoted as $\gamma$). The histogram (gray bars) is fit well by two Gaussian distributions plotted as two curves in the figure.}
\end{figure}

The layer-based matching does not consider the lateral or tangential span of neuron branches, which can also be an important feature of medulla neurons. Indeed, the histogram of the ratio of tangential span to columnar span shows an exponential distribution with a heavy tail, indicating a mixture of distributions. Converting the histogram into log-histogram shows more clearly that the distribution is close to the mixture of two log-normal distributions, which can also be estimated automatically by a decomposition algorithm \cite{figueiredo2002unsupervised} (Figure \ref{fig:lvratio}).

More formally, the distribution can be written as
\begin{equation}
f(x) = \sum_{i=1}^2 p_i f_i(x)
\label{eq:mixture}
\end{equation}
where
\begin{equation}
f_i(x) = \frac{1}{\sqrt{2\pi}\sigma_i}e^{-\frac{\ln(x)-\mu_i}{2\sigma_i^2}}
\end{equation}
is the normal distribution.

Assuming the prior fraction $p_i$ is evenly distributed, we can show that the probability that two neurons belong to the same class is

\begin{equation}
k(x, y) = \frac{r(x)r(y) + 1}{r(x)r(y) + r(x) + r(y) + 1}
\end{equation}
where $r(u) = \frac{f_1(u)}{f_2(u)}$.

The similarity score can be calibrated as

\begin{equation}
\tilde{S_{ij}} = \hat{S_{ij}}k^{\alpha}(\mathcal{N}_i, \mathcal{N}_j)
\end{equation}
where $\alpha$ is the parameter to control how much the tangential-columnar ratio feature contributes to the overall score. It is preferred to be less than 1 because such a value means that the features contributes more than linear scaling when the two neurons are less likely to belong to the same class. We found that the matching always became more accurate or stayed the same when $\alpha \le 0.5$.

\subsection{Workflow of Cell Type Identification}

\begin{figure}
\centering
\includegraphics[width=0.8\textwidth]{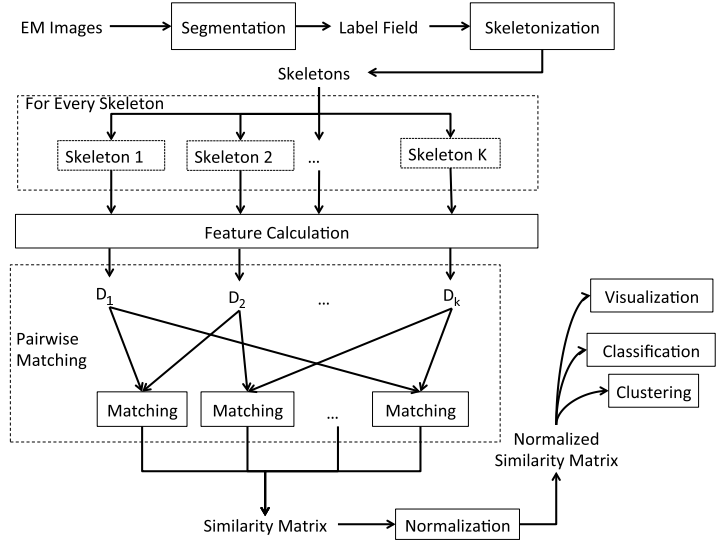}
\caption{\label{fig:flowchart} The workflow of cell type identification.}
\end{figure}

We incorporated neuron shape comparison into the pipeline of our neuron reconstruction (Figure \ref{fig:flowchart}), which starts from raw EM images as the input. The module of shape comparison generates a similarity matrix for neuron type identification and further statistical analysis. The details of some important steps in the workflow are described as follows.

\subsubsection{Neuron Reconstruction}

Our data, which corresponds to a sample of {\it Drosophila} optic medulla,  were originally acquired from electron microscopy. 379 neurons were reconstructed with a semi-automated approach and then the type of each neuron was identified manually by biological experts. Since these neurons are all from the same sample, along with the manual type labels as ground truth, they form an ideal set of data for testing our location-based neuron type identification method. After reconstruction, the neurons are represented in a 3D label volume, and to facilitate further analysis, we extracted the skeleton of each neuron by a robust skeletonization method.

\subsubsection{Skeletonization}

Skeletonization is the process of converting a binary image into a skeleton model. There are multiple ways of skeletonizing a binary object and we found the method called TEASAR \cite{sato2000teasar} to be most appropriate to our problem. In our implementation, we adopted TEASAR by applying the following steps:

\begin{enumerate}
\item Compute a 2D distance map on the foreground of each slice and denote it as $\mathcal{D}(\mathbf{v})$ , where $\mathbf{v} = (x, y, z)$  is the coordinates of a voxel. Let $\mathbf{v}_0 = \argmax_{\mathbf{v}} D(\mathbf{v})$   be the point with the maximal distance value.

\item Assuming all the foreground voxels form a graph in which each voxel is only connected to its direct neighbors (6-neighborhood system), compute the shortest geodesic paths from all voxels to $\mathbf{v}_0$ , with the geodesic distance between two neighbor points $\mathbf{v}_1$ and $\mathbf{v}_2$ defined as

\begin{equation}
d_g = \|\mathbf{v}_1 - \mathbf{v}_2\|_2 \left(\frac{1}{\alpha + I(\mathbf{v}_1)} + \frac{1}{\alpha + I(\mathbf{v}_2)}\right)
\end{equation}

where $I(\mathbf{v})$ is the image intensity at $\mathbf{v}$ and $\alpha$ is the regularization parameter, which is set to 1 in our implementation.

\item Find the farthest point to $\mathbf{v}_0$ in terms of the geodesic distance and initialize the source set $\mathcal{S} = \{B[\mathbf{v}'_0, \mathcal{D}(\mathbf{v}'_0)]\}$, where $B[\mathbf{v}, r]$ contains all voxels that are within the ball with center $\mathbf{v}$ and radius $r$.

\item Letting $\mathcal{T}$ be set of foreground voxels excluding $\mathcal{S}$, find the point $\mathbf{v} \in \mathcal{T}$ that has the largest distance to $\mathcal{S}$. Assuming the path between $\mathbf{v}$ and $\mathcal{S}$ is $(\mathbf{v}_1, ..., \mathbf{v}_n)$, if the path length is bigger than the threshold, let $\mathcal{S} = \mathcal{S} \cup \left(\cup_{i=1}^n B(\mathbf{v}_i, \mathcal{D}(\mathbf{v}_i))\right)$. Repeat step 5 until there is no valid path available.

\end{enumerate}

When there are multiple objects, or connected components, belonging to the same neuron, we skeletonize each object separately and then connect them in the order of building a minimal spanning tree to minimize the sum of connection distance.  This is the same as maximizing the skeleton likelihood if we assume that the objects that are closer to each other are more likely to be connected directly \cite{zhao2011automated}.

\subsubsection{Clustering}

The similarity matrix obtained from shape matching provides a good start for clustering because many clustering algorithms take a similarity/distance matrix directly as input and assign clustering labels to each sample. One of the most robust methods to this problem is affinity propagation (AP) \cite{frey2007clustering}, which finds clusters by measuring affinity among data points. It also has the advantage of identifying the number of clusters automatically by maximizing the net similarity under certain constraints. The AP clustering finds the most representative data sample called an examplar, which is a neuron in our case, for each cluster. This helps us evaluate the quality of clustering in a more interpretive way than using conventional consistency scores.

Another usage of similarity matrix is to project neurons into a low dimensional space and produce visualization of neuron clusters. A straightforward choice is Laplacian eigenmap (LE) \cite{belkin2003laplacian}, which generates low dimensional coordinates for each point to preserve the distances defined in the similarity matrix.

\subsubsection{Supervised Cell Type Classification}

Suppose we have a training set
consisting of a pool of
identified cells, then given a unidentified cell, we want to identify the cell by matching it to the cells in the training set. Since we have computed the matching scores or similarities, we can assign the dominating label of its neighbors to the cell using the kNN classifier. In our data, the number of neurons in each types varies from 1 to 37, and most types have no more than five samples. Therefore, we set $k=1$, which represents the nearest-neighbor classifier (NNC). With the similarity matrix, we can evaluate its quality by applying NCC on each neuron and summarizing the predication accuracies. 

\subsubsection{Implementation}

Neuron skeletonization and shape comparison were implemented in C++ as a part of NeuTu (https://github.com/janelia-flyem/NeuTu), our neuron reconstruction and visualization software. The Laplacian eigenmap was implemented in Matlab (The Mathworks, Inc.). The code of AP clustering was downloaded from http://www.psi.toronto.edu/affinitypropagation/software/apcluster.m and we used default parameters specified in the program in our application.

\section{Results}

\subsection{Data}

\begin{figure}
\centering
\includegraphics[width=1.0\textwidth]{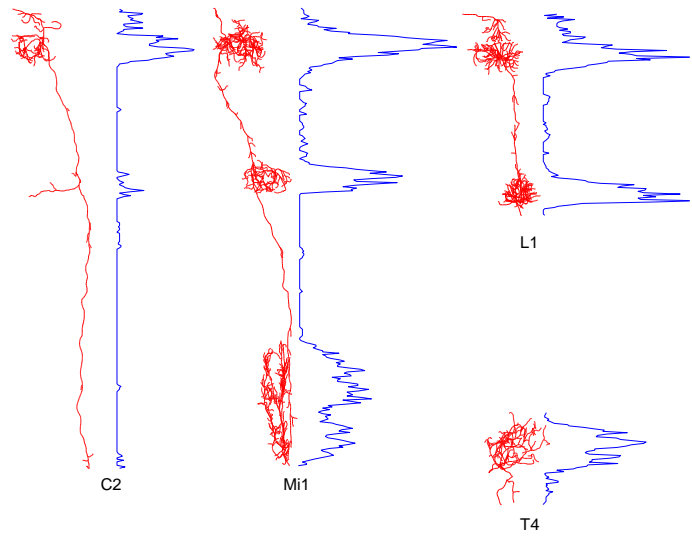}
\caption{\label{fig:layer_feature} Branch densities (blue) of several neurons (red) along the columnar direction. The relative columnar position of the neurons are preserved in the plot.}
\end{figure}

\begin{figure}
  \centering
    \begin{subfigure}[b]{0.3\textwidth}
     \includegraphics[width=\textwidth]{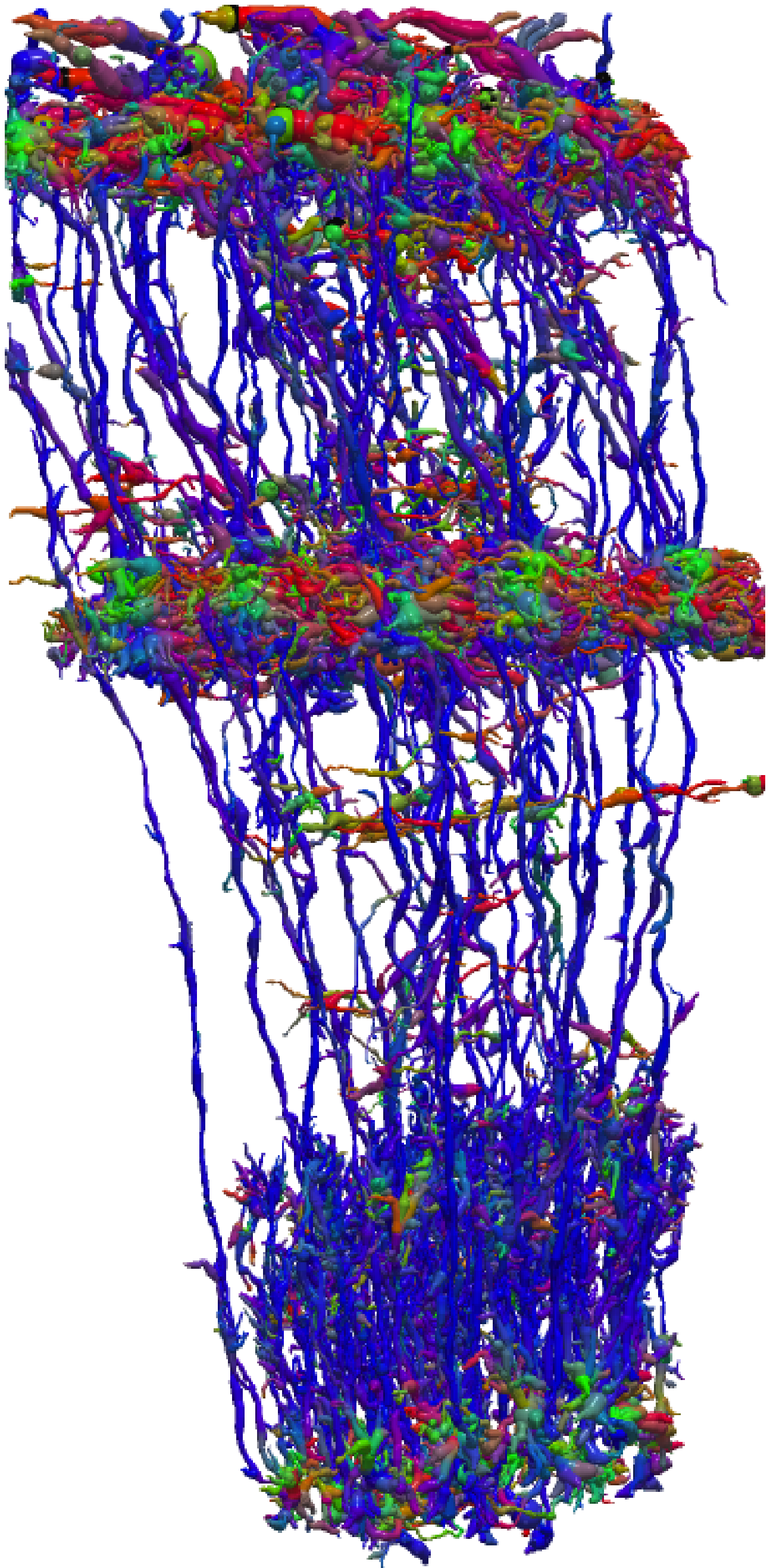}
    \caption{}
  \end{subfigure}
  \begin{subfigure}[b]{0.3\textwidth}
   \includegraphics[width=\textwidth]{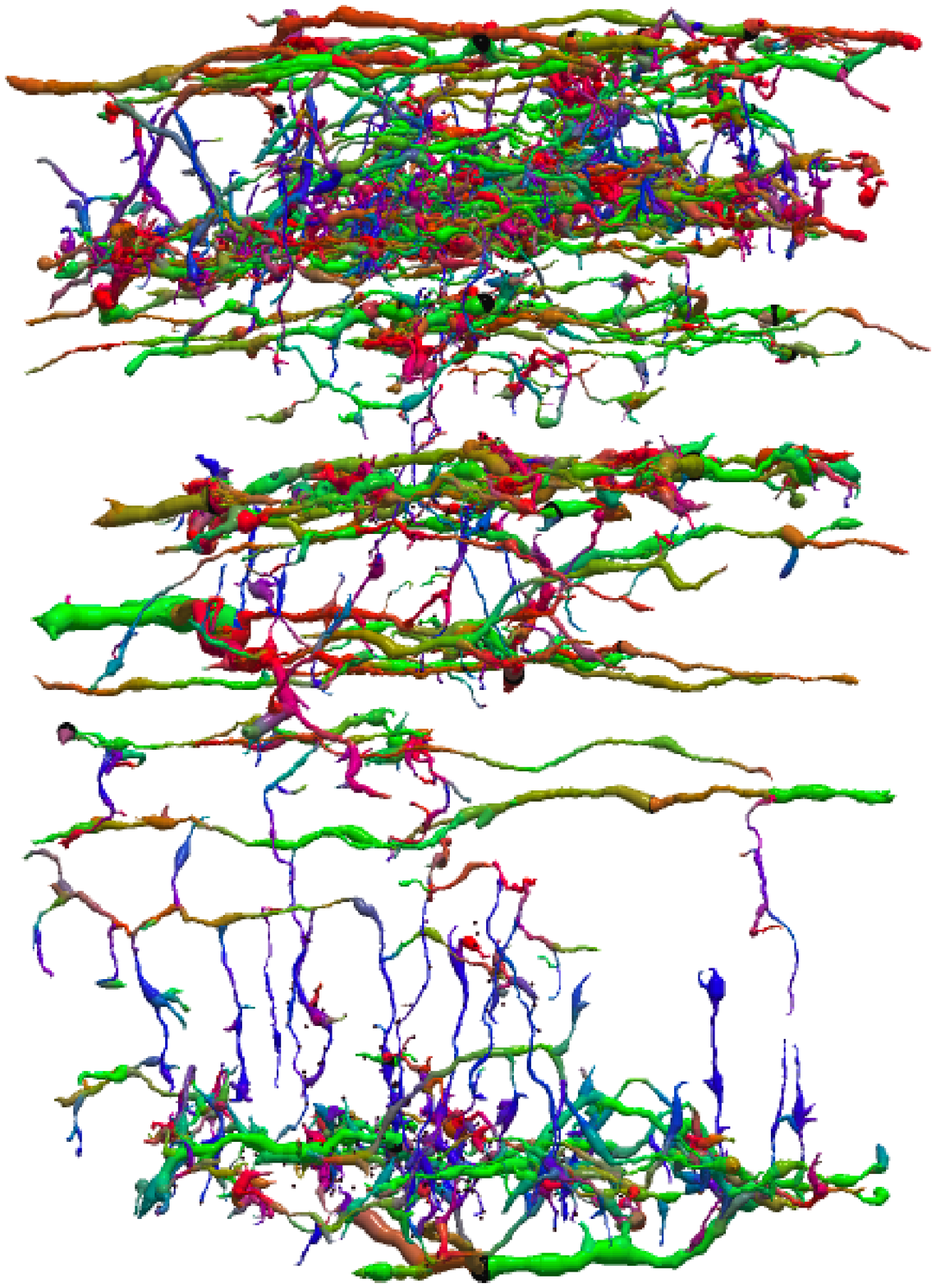}
    \caption{}
  \end{subfigure}
  \caption{\label{fig:columnar} Columnar neurons (a) and tangential neurons (b) visualized with color coding of branch directions (X: Red; Y: Green; Z: Blue).}
\end{figure}

The testing data contain 379 neurons that were manually categorized into 89 types. All the skeletons including the type information can be downloaded from http://neuromorpho.org under the `Drosophila - Chklovskii' category.  Figure \ref{fig:layer_feature} shows examples of several types and their branch distribution along the columnar direction. Another informative feature of these neurons is the tangential-columnar branch ratio, which is  automatically separated into two populations corresponding to neurons aligned with the columnar and tangential directions (Figure \ref{fig:columnar}).

\subsection{Classification Accuracy}

\begin{figure}
\centering
\includegraphics[height=0.9\textheight]{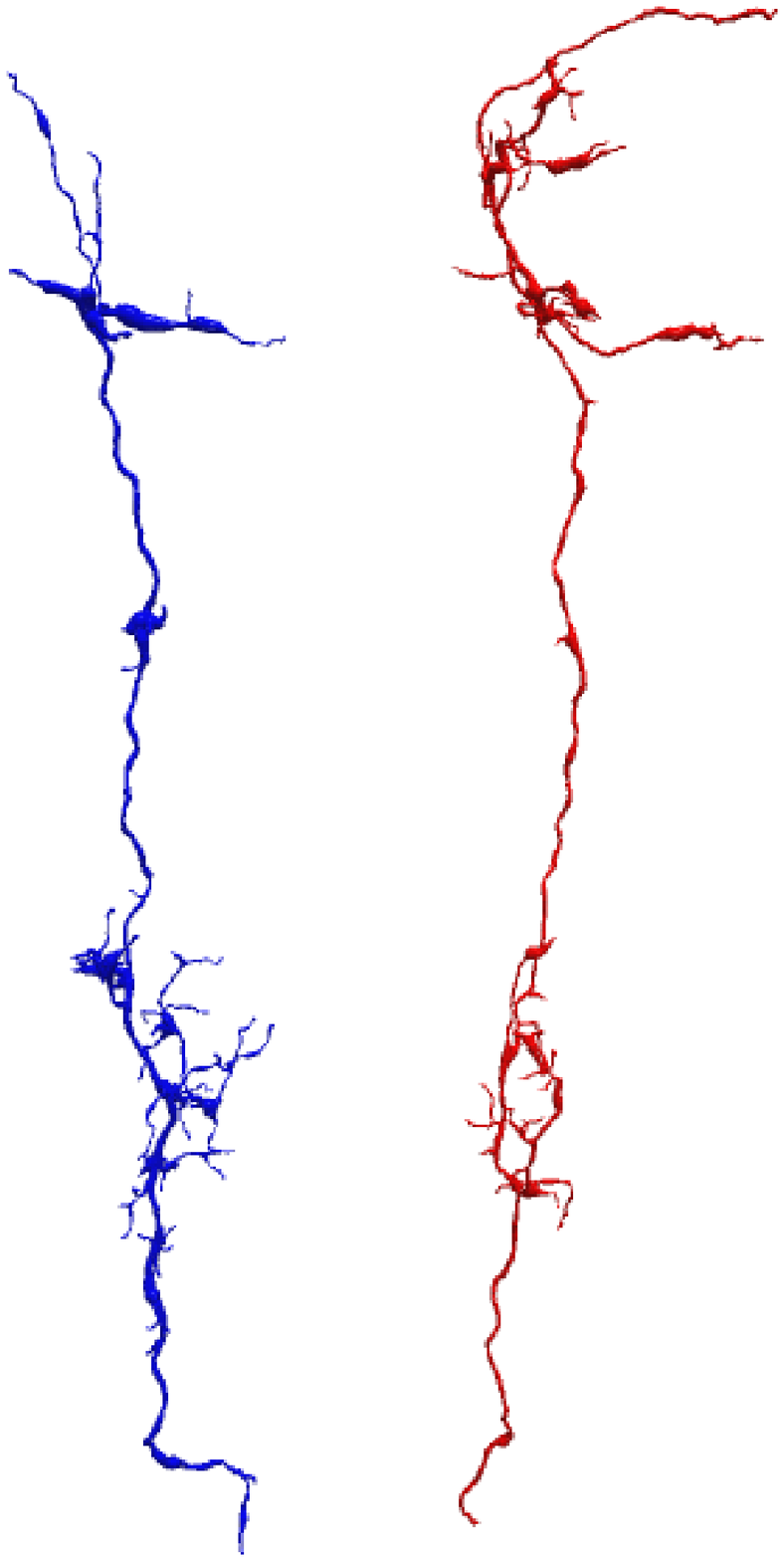}
\caption{\label{fig:T2_Tm6} A T2 neuron (a) is incorrectly matched to a Tm6 neuron (b). These two neurons are visually similar.}
\end{figure}

Among the 89 manually defined types, 33 types have only one sample, thereby leaving 346 neurons for testing classification accuracy. The overall accuracy of classifying these neurons with leave-one-out cross validation is 91\% (315/346). Without tangential calibration, the classification accuracy is slightly lower (313/346).
The details of classification errors are shown in Table \ref{table:classif}. Figure \ref{fig:T2_Tm6} shows an example of a mis-classification, where a T2 neuron is classified as the type Tm6 because it is most similar.

\begin{table}
\centering
\begin{tabular} {lcc}
Type&\#Mis-classification&\#Neurons\\
\hline
Dm10&1&2\\
Dm2&1&5\\
Dm3-like&1&3\\
Dm5-like&1&5\\
Mi1&2&19\\
Mi4&1&7\\
Tangential&12&37\\
Pm1&2&7\\
Pm2-like&2&12\\
T2&1&5\\
T2a&1&6\\
Tm16&1&2\\
unknown Tm-17&1&2\\
Tm5c-like&1&2\\
TmY10&1&2\\
TmY5-like&2&2\\
\hline
\end{tabular}
\caption{\label{table:classif} The list of neurons that are mis-classified.}
\end{table}

\subsection{Clustering Quality}


The AP clustering produced 51 clusters, of which the {\em examplar} neurons, or the most representative neurons, belong to 39 manually defined types (Table \ref{table:cluster}). By labeling each cluster as  the type of its examplar, we could compute the precision and recall of the cluster to evaluate how meaningful it is. Table \ref{table:cluster} shows that most of the clusters (42/51) have either precision or recall no lower than 80\% and 17 of them, including L1, Mi1 and Tm3, the neurons that are important to our previous motion circuit analysis, have both high precision and high recall. All of the clusters have either precision and recall higher than 50\%. The average precision and recall, which are computed by averaging the precision and recall values with equal weight of each cluster, are 74.5\% and 72.3\% respectively. The nine clusters that are labeled as the Tangential type have relatively high precision but low recall, indicating that the tangential neurons are separated into different clusters. This is expected because this type contains a superclass of tangential neurons whose named types cannot be identified \cite{takemura2013visual}. The recall is 81.3\% if we merge these clusters into one, increasing the average recall to 86.0\%. For those types without examplars, which accounts for 21.9\% of the neuron population, the numbers of samples are small, with most of them having only one sample. Therefore, the clustering algorithm tends to put them in a bigger cluster. 

\begin{table}
\centering
\begin{tabular}{llrr}
Cluster& Examplar Type & Precision & Recall\\
\hline
1&\textcolor{black}{C2} & 62.5\% & 100\%\\
2&\colorbox{green}{C3} & 100\% & 100\%\\
3&\textcolor{black}{Dm1-like} & 25\% & 100\%\\
4&\textcolor{black}{Dm2} & 66.7\% & 80\%\\
5&\textcolor{black}{Dm4} & 66.7\% & 100\%\\
6&\textcolor{red}{Dm5-like} & 42.9\% & 60\%\\
7&\colorbox{green}{Dm8} & 80\% & 80\%\\
8&\colorbox{green}{L1} & 100\% & 94.7\%\\
9&\textcolor{black}{L2} & 70\% & 100\%\\
10&\colorbox{green}{L3} & 87.5\% & 100\%\\
11&\colorbox{green}{L4} & 100\% & 100\%\\
12&\colorbox{green}{L5} & 87.5\% & 100\%\\
13&\colorbox{green}{Mi15} & 80\% & 100\%\\
14&\colorbox{green}{Mi1} & 100\% & 89.5\%\\
15&\colorbox{green}{Mi4} & 100\% & 85.7\%\\
16&\textcolor{black}{Mi9} & 77.8\% & 100\%\\
17&\textcolor{black}{Mt11-like} & 33.3\% & 100\%\\
18&\textcolor{red}{Pm1} & 66.7\% & 28.6\%\\
19&\textcolor{red}{Pm1} & 60\% & 42.9\%\\
20&\textcolor{red}{Pm2-like} & 57.1\% & 33.3\%\\
21&\textcolor{red}{Pm2-like} & 75\% & 50\%\\
22&\textcolor{black}{unknown Pm-1} & 36.4\% & 100\%\\
23&\colorbox{green}{R7} & 90\% & 100\%\\
24&\colorbox{green}{R8} & 100\% & 100\%\\
25&\textcolor{black}{T3} & 75\% & 100\%\\
26&\textcolor{black}{T4} & 100\% & 41.4\%\\
27&\textcolor{black}{T4} & 100\% & 24.1\%\\
28&\colorbox{green}{Tm1} & 100\% & 100\%\\
29&\colorbox{green}{Tm20} & 100\% & 100\%\\
30&\colorbox{green}{Tm2} & 90.9\% & 100\%\\
31&\colorbox{green}{Tm3} & 86.4\% & 90.5\%\\
32&\textcolor{black}{Tm4} & 63.6\% & 100\%\\
33&\textcolor{black}{Tm5b} & 7.7\% & 100\%\\
34&\textcolor{black}{Tm6/14} & 30.8\% & 100\%\\
35&\textcolor{black}{Tm9} & 33.3\% & 100\%\\
36&\colorbox{green}{TmY3} & 80\% & 100\%\\
37&\textcolor{black}{TmY5a} & 57.1\% & 80\%\\
38&\textcolor{black}{Y4} & 66.7\% & 100\%\\
39&\colorbox{green}{Y3/Y6} & 100\% & 100\%\\
40&\textcolor{black}{T4} & 100\% & 34.5\%\\
41&\textcolor{black}{unknown Tm-15} & 5.9\% & 100\%\\
42&\textcolor{black}{LaWF1} & 66.7\% & 100\%\\
43-51&\textcolor{red}{Tangential}* & 62.5\% - 100\% & 5.4\% - 13.5\%\\
\hline
\end{tabular}
\caption{\label{table:cluster} The quality of clusters produced by AP clustering. 17 clusters (green background) have both high precision and high recall ($\ge 80\%$ ) and nine clusters (red text), including four Tangential clusters, have no high precision or recall.\\
*\footnotesize{The Tangential type is separated into nine clusters. Merging these clusters into one gives a cluster with 70.3\% precision and 81.3\% recall.}
}
\end{table}



\begin{figure}
\centering
\includegraphics[width=0.8\textwidth]{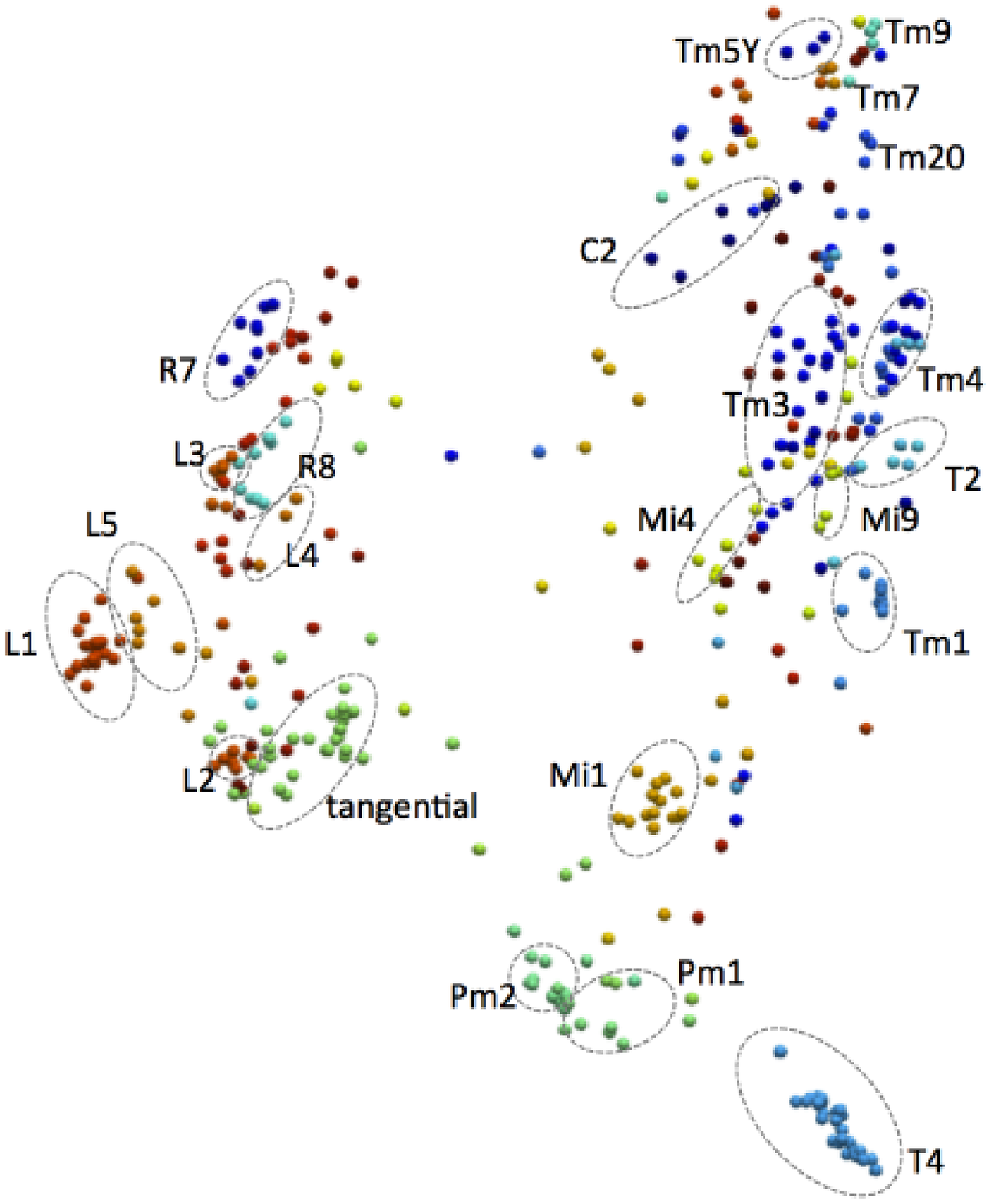}
\caption{The LE projection of medulla neurons visualizes clusters of neurons, each rendered as a ball with a color indicating its type. }
\label{fig:le_proj} 
\end{figure}

To facilitate visually checking the clustering patterns, we projected the neurons to a three dimensional space using the Laplacian eigenmap (LE) method. The clusters are well preserved after the projection, revealing an interesting pattern of two main arms of the point cloud (Figure  \ref{fig:le_proj}), which correspond to two neuron populations related to their processing depth defined in \cite{takemura2013visual}.


\subsection{Discussion}

We presented a method that is very different than traditional neuron clustering methods based on shape features. Our method introduces and tests the idea that the types of neurons are mainly determined by where their branches are located. The high accuracy of identifying neuron types by the branch density along layers only suggests that where the branches project can be more important than how the branches project for neuron functions. 

Although our method requires all the neurons to be aligned in the same framework, it can be generalized to a broader problem domain as long as the alignment can be reasonably defined, such as aligning sparsely labeled neurons in the same standard brain \cite{peng2011brainaligner}. Note that Eq. \ref{eq:matchg} does not require reconstruction of neuron structures, which is a much more difficult problem for low-resolution microscopy, making the method well suited for matching neurons across imaging modalities. For example, we can simply define the branch density as average intensity over a certain region for a light microscope image. 

Our method automatically identified neuron types with a sufficient number of samples. No predefined number of clusters is necessary. Although the neurons in medulla have been well studied and their types have been well-defined, this is the first time that we generated those types in an objective way, confirming that there are morphologically distinguishable groups among those cells.

\begin{figure}
\centering
\includegraphics[width=0.8\textwidth]{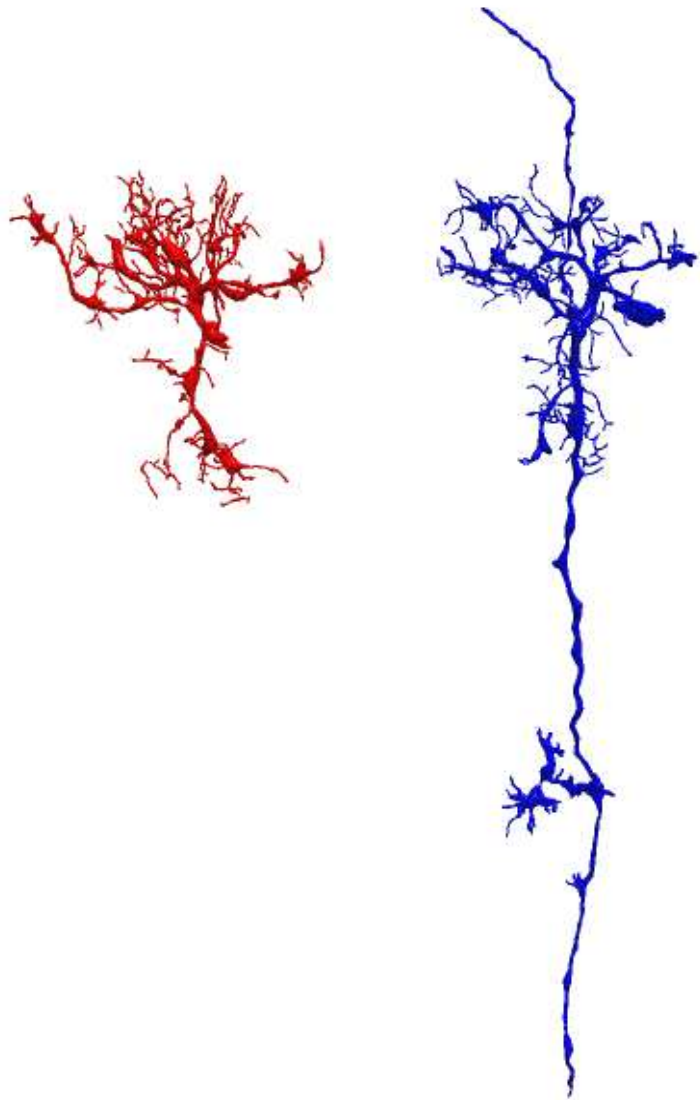}
\caption{A partial Tm4 neuron (a) is clustered together with a complete Tm4 neuron (b).}
\label{fig:partialmatch} 
\end{figure}

Automatic neuron matching and type identification provides not only  new biological insights, but also quality control of connectome analysis. It can be used to identify partially reconstructed neurons as in Figure \ref{fig:partialmatch}. We have used it to help biologists label cells in ongoing reconstructions as well as to find partially reconstructed neurons. In the medulla reconstructions, our clustering has confirmed that some cell types are missing in some columns.

\bibliography{reference.bbl}

\end{document}